\documentclass[journal]{IEEEtran}

\usepackage{subfigure}
\usepackage{setspace}
\usepackage{multirow} 
\usepackage{algorithm}
\usepackage{algorithmic}
\usepackage{amsmath}
\usepackage{amssymb}
\usepackage{latexsym}
\usepackage{multirow}
\usepackage{epsfig}
\usepackage{graphics}
\usepackage{graphicx}
\usepackage{mathrsfs}
\usepackage{subfigure}
\usepackage{slashbox}
\usepackage{mathrsfs}
\usepackage{bbding}
\usepackage{cite}
\usepackage{color}
\usepackage{xcolor}
\usepackage{booktabs}
\usepackage{amssymb,mathrsfs,amsmath}
\newcommand{\bm}[1]{\mbox{\boldmath{$#1$}}}
\usepackage{cases}
\usepackage{cancel}
\usepackage{times}
\usepackage{multirow}
\usepackage{booktabs}

\usepackage{hyperref}

\usepackage{amsthm}
\theoremstyle{plain}

\allowdisplaybreaks [4]

\begin{document}

\title{IRS-Aided Multi-Antenna Wireless Powered Communications in Interference Channels}
\author{Ying~Gao, Qingqing~Wu,~\IEEEmembership{Senior Member,~IEEE}, and Wen~Chen,~\IEEEmembership{Senior Member,~IEEE} \vspace{-4mm}
\thanks{The work of Qingqing Wu was supported by National Key R\&D Program of China (2023YFB2905000), NSFC 62371289, and NSFC 62331022. The work of Wen~Chen was supported by National key project NSFC 62071296 and Shanghai Kewei 22JC1404000. The authors are with the Department of Electronic Engineering, Shanghai Jiao Tong University, Shanghai 201210, China (e-mail: yinggao@sjtu.edu.cn; qingqingwu@sjtu.edu.cn; whenchen@sjtu.edu.cn).}}

\maketitle

\begin{abstract}
	This paper investigates intelligent reflecting surface (IRS)-aided multi-antenna wireless powered communications in a multi-link interference channel, where multiple IRSs are deployed to enhance the downlink/uplink communications between each pair of hybrid access point (HAP) and wireless device. Our objective is to maximize the system sum throughput by optimizing the allocation of communication resources. To attain this objective and meanwhile balance the performance-cost tradeoff, we propose three transmission schemes: the IRS-aided asynchronous (Asy) scheme, the IRS-aided time-division multiple access (TDMA) scheme, and the IRS-aided synchronous (Syn) scheme. For the resulting three non-convex design problems, we propose a general algorithmic framework capable of suboptimally addressing all of them. Numerical results show that our proposed IRS-aided schemes noticeably surpass their counterparts without IRSs in both system sum throughput and total transmission energy consumption at the HAPs. Moreover, although the IRS-aided Asy scheme consistently achieves the highest sum throughput, the IRS-aided TDMA scheme is more appealing in scenarios with substantial cross-link interference and limited IRS elements, while the IRS-aided Syn scheme is preferable in low cross-link interference scenarios.
\end{abstract}


\vspace{-4mm}
\section{Introduction}
As a typical application of wireless power transfer (WPT), wireless powered communication network (WPCN) is viewed as a promising network paradigm to address the energy shortage problem of wireless devices (WDs). In \cite{2014_Hyungsik_WPCN}, the authors studied a single-input single-output (SISO) WPCN and proposed the well-known ``harvest-then-transmit'' protocol. This work was then extended to more general single-cell multi-antenna WPCNs in \cite{2014_Liang_WPCN} and \cite{2016_Hoon_WPCN}, and a multi-cell SISO WPCN in \cite{2018_Hanjin_WPCN}, respectively. Despite theoretical progress, practical WPCNs face severe limitations due to the low efficiencies of WPT over long transmission distances. Recently, intelligent reflecting surface (IRS) \cite{2019_Qingqing_Joint,2021_Xiaowei_IRS,2023_Yujie_Covert} has emerged as an effective remedy to tackle this performance bottleneck. The results in \cite{2021_Qingqing_NOMA} indicated that introducing an IRS into a SISO WPCN has the dual benefits of enhancing the system throughput and lowering the energy consumption at the hybrid access point (HAP). Additionally, the authors of \cite{2021_Xingquan_WPCN} and \cite{2022_Hua_WPCN} confirmed that the IRS can improve the performance of multi-antenna WPCNs. \looseness=-1 

However, all the above-mentioned works on IRS-assisted WPCNs are limited to single-cell scenarios. To the best of the authors' knowledge, the potential performance gain of integrating IRSs into multi-cell WPCNs in interference channels (IFCs) has not been explored yet. Recall that the authors of \cite{2018_Hanjin_WPCN} studied a traditional multi-cell SISO WPCN. Their simulation results demonstrated that an asynchronous protocol involving asynchronous downlink (DL)-uplink (UL) time allocation for different cells can lead to a larger system sum throughput compared to its synchronous counterpart. Intuitively, integrating IRSs into multi-cell WPCNs can enhance system performance by strategically designing IRS phase shifts in each time slot to customize favorable time-varying channels. However, formulating the corresponding problem and designing the resource allocation algorithm are notably dissimilar and markedly more arduous compared to those in \cite{2018_Hanjin_WPCN}. This is primarily due to the introduction of additional time-varying IRS phase-shift optimization variables, rendering the algorithm proposed in \cite{2018_Hanjin_WPCN} inapplicable. Moreover, the work in \cite{2018_Hanjin_WPCN} has the following limitations. First, it was confined to a SISO setting without harnessing the multi-antenna technology to improve the system performance. Second, it did not disclose whether the performance gain achieved by the asynchronous protocol comes with higher energy costs, a factor that influences the selection of transmission protocols in practice. 

\begin{figure}[!t]
	\vspace{-2mm}
	\hspace{-3mm}
	\centering
	\subfigbottomskip = -0.1pt  
	\subfigure[]{\label{fig:asy}
		\includegraphics[width = 0.158\textwidth]{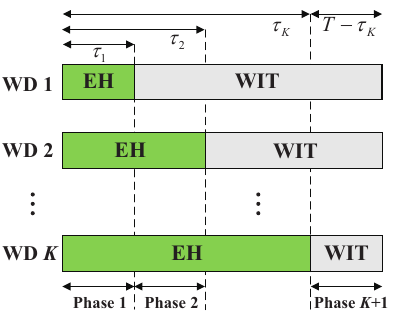}}
	\hspace{-3mm}
	\subfigure[]{\label{fig:TDMA}
		\includegraphics[width = 0.158\textwidth]{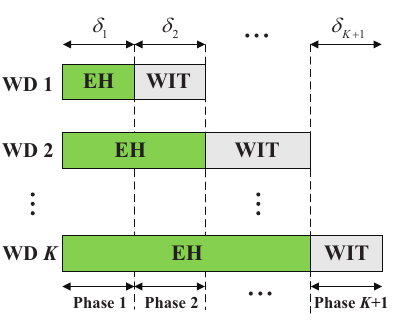}}
	\hspace{-3mm}	
	\subfigure[]{\label{fig:sy}
		\includegraphics[width = 0.158\textwidth]{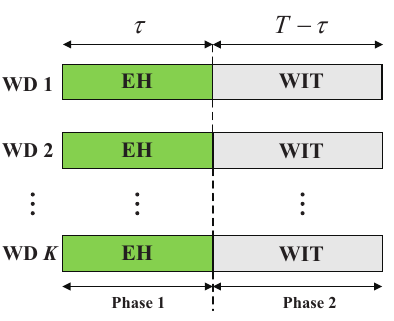}} 
	\caption{Illustration of three different transmission schemes: (a) IRS-aided Asy scheme; (b) IRS-aided TDMA scheme; (c) IRS-aided Syn scheme.}
	\vspace{-6mm}
\end{figure}

Motivated by the above considerations, this paper studies an IRS-aided multi-antenna wireless powered IFC, comprising multiple IRSs and HAP-WD pairs. To maximize the system sum throughput while considering the balance between performance and cost, we propose three transmission schemes: the IRS-aided asynchronous (Asy) scheme, the IRS-aided time-division multiple access (TDMA) scheme, and the IRS-aided synchronous (Syn) scheme, ranked from high to low in terms of implementation complexity. Theoretically, the IRS-aided Asy scheme always attains the highest sum throughput, as it is a super-scheme of the other two. In addition, neither the IRS-aided TDMA scheme nor the IRS-aided Syn scheme can consistently outperform the other in terms of system sum throughput. This is mainly because the IRS-aided TDMA scheme enjoys the advantage of being free from cross-channel interference in uplink wireless information transfer (WIT) but grapples with the inefficient utilization of time resources. This scheme's advantage tends to diminish or even disappear in scenarios with low cross-link interference and/or sufficient IRS elements in the IRS-aided Syn scheme, while its drawback remains challenging to overcome even with an increased number of IRS elements. Nevertheless, theoretically ranking the energy consumption of these three schemes is difficult. We propose a general algorithmic framework to solve the three corresponding resource allocation problems suboptimally and numerically evaluate the performance of the three schemes. Simulation results verify the above discussions regarding performance comparisons and show that the IRS-aided TDMA scheme consumes the most energy while the IRS-aided Syn scheme consumes the least. After a comprehensive evaluation of performance, implementation complexity, and energy cost, we find that the IRS-aided TDMA scheme is most appealing in scenarios with overwhelming cross-link interference and limited IRS elements, the IRS-aided Syn scheme is the best choice when cross-link interference is low, and the IRS-aided Asy scheme is preferable in all other scenarios.

\emph{Notations:} $(\cdot)^T$, $(\cdot)^*$, $(\cdot)^H$, and ${\rm tr}(\cdot)$ denote the transpose, conjugate transpose, Hermitian transpose, and trace operators, respectively. $\left|\cdot\right|$ stands for the cardinality of a set or the modulus of a complex number. $\left\|\cdot\right\|$ denotes the Euclidean norm of a vector. ${\rm diag}(\mathbf a)$ creates a diagonal matrix with each diagonal element being the corresponding element in vector $\mathbf a$. ${\rm Diag}(\mathbf A)$ forms a vector with elements extracted from the diagonal of matrix $\mathbf A$. ${\rm blkdiag}(\mathbf A_1,\cdots,\mathbf A_N)$ generates a block diagonal matrix with diagonal components $\mathbf A_1,\cdots,\mathbf A_N$. 

\vspace{-3mm}
\section{System Model and Problem Formulation}\label{Sec_model_formu}
\subsection{System Model}
We consider a narrow-band IRS-aided wireless powered IFC, where $K$ pairs of $M$-antenna HAPs and single-antenna WDs operate over the same frequency band with the aid of $L$ passive IRSs \footnote{In this initial study on multi-IRS-aided multi-cell WPCNs, we focus on the basic passive IRS architecture. We note that research \cite{2022_Kangda_activeRIS} suggested that with equal power budgets, an active IRS outperforms a passive one in terms of achievable system rate under certain setups, due to its ability to mitigate double-fading attenuation. While our proposed algorithmic framework remains applicable for active IRSs, specific adjustments are necessary to accommodate the newly imposed amplification power constraints and the non-negligible IRS-amplified noise power. We leave this and the performance comparison between active and passive IRSs-aided multi-cell WPCNs for future study.}. The $\ell$-th IRS is equipped with $N_{\ell}$ elements. The sets of HAP-WD pairs, IRSs, and all the IRS elements are denoted as $\mathcal K$, $\mathcal L$, and $\mathcal N$, respectively, with $\left|\mathcal K\right|= K$, $\left|\mathcal L\right| = L$, and $\left| \mathcal N\right| = N = \sum_{\ell = 1}^LN_{\ell}$. We assume that the the HAPs/IRSs possess accurate knowledge of the channel state information (CSI) of all channels involved, which can be obtained through active sensors mounted on the IRSs or by using existing channel estimation methods discussed in \cite{2022_Cunhua_overview}. The UL channels from WD $k$ to IRS $\ell$, from IRS $\ell$ to HAP $i$, and from WD $k$ to HAP $i$ are denoted by $\mathbf e_{k,\ell} \in\mathbb C^{N_\ell\times 1}$, $\mathbf H_{\ell,i}\in\mathbb C^{M\times N_\ell}$, and $\mathbf g_{k,i} \in \mathbb C^{M\times 1}$, respectively, which remain constant throughout the total transmission time $T$. Moreover, the UL-DL channel reciprocity is assumed. 

\subsubsection{IRS-Aided Asy Scheme}\label{Subsec:asy}
Following the typical ``harvest-then-transmit'' protocol \cite{2014_Hyungsik_WPCN}, WD $k$ performs EH first and then WIT, with durations of $\tau_k$ and $T-\tau_k$, respectively. Without loss of generality, we assume that $\{\tau_k\}$ are sorted in an ascending order, i.e., $\tau_1 \leq \tau_2\leq\dots\leq\tau_K$. According to the WDs' operations, the entire transmission process can be divided into $K+1$ phases, as illustrated in Fig. \ref{fig:asy}. We define $\delta_j \triangleq \tau_j - \tau_{j-1}$ as the time duration of phase $j$ ($j\in\mathcal J \triangleq \{1,\cdots,K+1\}$), where $\tau_0 \triangleq 0$ and $\tau_{K+1} \triangleq T$. Denoted by $\mathbf \Theta_{\ell,j} = {\rm diag}\left(\beta_{\ell,1,j}e^{\jmath\theta_{\ell,1,j}}, \cdots, \beta_{\ell,N_\ell,j}e^{\jmath\theta_{\ell,N_\ell,j}} \right)$ the reflection-coefficient matrix at IRS $\ell$ in phase $j$, where $\beta_{\ell,n,j}$ and $\theta_{\ell,n,j}$ can be independently adjusted over $[0,1]$ and $[0,2\pi)$, respectively \cite{2016_Huanhuan_amplitude}. Besides, let $\bm x_{i,j} \in\mathbb C^{M\times 1}$ denote the energy signal vector transmitted by HAP $i$ in phase $j$ ($j = 1,\cdots,i$), with a covariance matrix $\mathbf S_{i,j} = \mathbb E\left(\bm x_{i,j}\bm x_{i,j}^H\right) \succeq \mathbf 0$. 

In phase $1$, all the HAPs are engaged in WPT. The received signal at WD $k$, $k\in\mathcal K$, in this phase is given by $y_{k,1}^{\rm E}  = \sum_{i=1}^K\left(\sum_{\ell=1}^L\mathbf H_{\ell,i}\mathbf \Theta_{\ell,1}\mathbf e_{k,\ell} + \mathbf g_{k,i}\right)^T\bm x_{i,1} + n_k = \sum_{i=1}^K\left(\mathbf H_i\mathbf \Theta_1\mathbf e_k + \mathbf g_{k,i} \right)^T\bm x_{i,1} + n_k = \sum_{i=1}^K\mathbf v_1^T\mathbf \Psi_{k,i}^T\bm x_{i,1} + n_k$, where $n_k$ stands for the additive white Gaussian noise (AWGN) at WD $k$, $\mathbf H_i = \left[\mathbf H_{1,i}, \cdots, \mathbf H_{L,i}\right]$, $\mathbf \Theta_1 = {\rm blkdiag}\left(\mathbf \Theta_{1,1}, \cdots, \mathbf \Theta_{L,1}\right)$, $\mathbf e_k = \left[\mathbf e_{k,1};\cdots;\mathbf e_{k,L}\right]$, $\mathbf v_1 = \left[{\rm Diag}\left(\mathbf \Theta_1\right);1\right]$ denotes the phase-shift vector associated with all the IRSs in phase $1$, and $\mathbf \Psi_{k,i} = \left[\mathbf H_i{\rm diag}\left(\mathbf e_k\right), \mathbf g_{k,i}\right]$. Note that the signals experiencing multiple reflections are disregarded due to the multiplicative path loss. By ignoring the negligible noise power, the energy harvested by MD $k$ in the 1st phase can be written as $E_{k,1} = \eta\delta_1\sum_{i = 1}^K{\rm tr}\left(\mathbf \Psi_{k,i}^*\mathbf v_1^*\mathbf v_1^T\mathbf \Psi_{k,i}^T\mathbf S_{i,1}\right)$, with $\eta$ denoting the energy conversion efficiency of each WD. 

In the subsequent phase $j$ ($j = 2, \cdots, K$), HAPs $\{i'|i' = j,\cdots,K\}$ continue to broadcast energy, whereas HAPs $\{i|i = 1, \cdots, j-1\}$ receive UL information signals from WDs $\{k|k = 1, \cdots, j-1\}$. The received signal at HAP $i$ in phase $j$ can be expressed as $\mathbf y_{i,j}^{\rm I} = \sum_{k=1}^{j-1}\left(\sum_{\ell=1}^L\mathbf H_{\ell,i}\mathbf \Theta_{\ell,j}\mathbf e_{k,\ell} + \mathbf g_{k,i} \right)\sqrt{p_{k,j}}s_k + \boldsymbol {\iota}^{\rm E}_{i,j} + \hat{\mathbf n}_i  = \sum_{k=1}^{j-1}\mathbf \Psi_{k,i}\mathbf v_j\sqrt{p_{k,j}}s_k + \boldsymbol {\iota}^{\rm E}_{i,j} + \hat{\mathbf n}_i$, where $p_{k,j}$ denotes the UL transmit power of WD $k$ in phase $j$, $s_k \in\mathbb C$ is the transmitted data symbol of WD $k$ satisfying $s_k\sim\mathcal{CN}\left( 0,1\right)$, $\boldsymbol {\iota}^{\rm E}_{i,j}$ stands for the DL-to-UL interference caused by the energy signals from HAPs $\{i'|i' = j,\cdots,K\}$, $\mathbf v_j = \left[{\rm Diag}\left(\mathbf \Theta_j\right);1\right]$ with $\mathbf \Theta_j = {\rm blkdiag}\left(\mathbf \Theta_{1,j}, \cdots, \mathbf \Theta_{L,j}\right)$, and $\hat{\mathbf n}_i \in\mathbb C^{M\times 1}$ represents the zero-mean AWGN at HAP $i$ with co-variance matrix $\sigma_i^2\mathbf I_M$. By assuming that the HAPs employ linear receivers to decode $\{s_k\}$, we denote $\mathbf w_{i,j} \in\mathbb C^{M\times 1}$ as the unit-norm receive beamforming vector at HAP $i$ for decoding $s_i$ in phase $j$. Additionally, we assume that the energy signals are known deterministic signals, enabling HAP $i$ to cancel the energy signal interference. Consequently, the achievable rate at HAP $i$ in bits/Hz during phase $j$ ($j = i+1, \cdots, K$) is obtained as $R_{i,j} = \delta_j\log_2\left(1 + \gamma_{i,j}\right)$ with $\gamma_{i,j} = \frac{p_{i,j}\left|\mathbf w_{i,j}^H\mathbf \Psi_{i,i}\mathbf v_j\right|^2}{\sum_{k=1,k\neq i}^{j-1}p_{k,j}\left|\mathbf w_{i,j}^H\mathbf \Psi_{k,i}\mathbf v_j\right|^2 + \sigma_i^2\left\|\mathbf w_{i,j} \right\|^2}$. Meanwhile, WDs $\{k|k = j,\cdots,K\}$ harvest energy in phase $j$ ($j = 2,\cdots, K$). For the EH at WD $k$, we ignore the UL WIT signal power from other WDs and the noise power, as both are negligible compared to the HAP transmit power. Then, the energy collected by WD $k$ in phase $j$ can be expressed as $E_{k,j} = \eta\delta_j\sum_{i = j}^K{\rm tr}\left(\mathbf \Psi_{k,i}^*\mathbf v_j^*\mathbf v_j^T\mathbf \Psi_{k,i}^T\mathbf S_{i,j}\right)$. 

In the final phase, all the WDs execute the WIT operation. The achievable rate at HAP $i$ is given by $R_{i,K+1} = \delta_{K+1}\log_2\left(1 + \gamma_{i,K+1} \right)$, where the expression of $\gamma_{i,K+1}$ can be obtained by replacing the index ``$j$'' in $\gamma_{i,j}$ by ``$K+1$''. 

\subsubsection{IRS-Aided TDMA Scheme}
For the scheme shown in Fig. \ref{fig:TDMA}, the WDs perform UL WIT in a TDMA manner. This scheme is a special case of the IRS-aided Asy scheme with $p_{k,j} = 0$ and $\mathbf w_{i,j} = \mathbf 0$, $\forall k,i\in\mathcal K, j \in\mathcal J\backslash \{k+1\}$.

\subsubsection{IRS-Aided Syn Scheme}
When $\tau_k = \tau$, $\forall k \in\mathcal K$, the IRS-aided Asy scheme boils down to the IRS-aided Syn scheme, as illustrated in Fig. \ref{fig:sy}. 

\vspace{-4mm}
\subsection{Problem Formulation}
For the IRS-aided Asy scheme, the sum throughput maximization problem can be formulated as
\begin{subequations}
	\begin{eqnarray}
	    \hspace{-4mm}\text{(P1)}: \hspace{-2mm}&\underset{\mathcal Z}{\max}& \sum_{i=1}^K\sum_{j=i+1}^{K+1}R_{i,j} \\
		&\text{s.t.}& \hspace{-3mm} \sum_{j=k+1}^{K+1}\delta_jp_{k,j} \leq \sum_{j=1}^kE_{k,j}, \forall k \in\mathcal K, \label{P1_cons:b}\\
		&& \hspace{-3mm} \sum_{j=1}^{K+1}\delta_j \leq T, \label{P1_cons:c}\\
		&& \hspace{-3mm}{\rm tr}\left(\mathbf S_{i,j}\right) \leq P_i, \forall i\in\mathcal K, j = 1,\cdots, i, \label{P1_cons:d}\\
		&& \hspace{-3mm} \left\|\mathbf w_{i,j}\right\|^2 = 1, \forall i\in\mathcal K, j = i+1,\cdots, K+1, \label{P1_cons:e}\\
		&& \hspace{-3mm} \left|\left[\mathbf v_j\right]_n\right| \leq 1, \ \left[ \mathbf v_j\right]_{N+1} = 1, \ \forall n\in\mathcal N, j \in\mathcal J,\label{P1_cons:f}
	\end{eqnarray}
\end{subequations} 
where $\mathcal Z\triangleq \left\lbrace \{\delta_j \geq 0\}, \{\mathbf S_{i,j} \succeq \mathbf 0\}, \{p_{k,j} \geq 0\}, \{\mathbf w_{i,j}\}, \{\mathbf v_j\}\right\rbrace$ is composed of all the optimization variables, including the time allocation, the DL energy variance matrices, the UL power allocation, the UL receive beamforming vectors, and the IRS reflect beamforming vectors. Furthermore, \eqref{P1_cons:b} and \eqref{P1_cons:c} denote the energy causality and total time constraints, respectively, $P_i$ in \eqref{P1_cons:d} represents the maximum instantaneous transmit power of HAP $k$, \eqref{P1_cons:e} means that $\{\mathbf w_{i,j}\}$ have unit norms, and \eqref{P1_cons:f} imposes modulus constraints on the IRS bemaforming vectors. 
Similarly, we can formulate the sum throughput maximization problems for the IRS-aided TDMA and Syn schemes, respectively, denoted by (P2) and (P3). The formulations of these two problems are omitted due to the space limitation. We note that the coupling of the optimization variables presents a challenge in solving (P1)-(P3) optimally. To this end, we propose a general algorithmic framework based on the alternating optimization technique, applicable to solving all of these problems suboptimally, as detailed in the following section. \looseness=-1 

\vspace{-2mm}
\section{Proposed General Algorithmic Framework for (P1)-(P3)}
\subsection{How To Solve (P1)?} 
Based on the principle of alternating optimization, we partition the optimization variables into three blocks, i.e., $\{\mathbf w_{i,j}\}$, $\left\lbrace \{\delta_j\}, \{\mathbf S_{i,j}\}, \{p_{k,j}\}\right\rbrace$, and $\{\mathbf v_j\}$. These blocks are updated alternately until convergence is achieved, as elaborated below. 

\subsubsection{Optimizing $\{\mathbf w_{i,j}\}$} 
When given other variables, the optimization of $\{\mathbf w_{i,j}\}$ can be performed independently and in parallel for each $\mathbf w_{i,j}$. Specifically, we define $\mathbf a_{k,i,j} \triangleq \mathbf \Psi_{k,i}\mathbf v_j$, $\forall k,i\in\mathcal K, j = i+1,\cdots, K+1$, and calculate $\mathbf w_{i,j}$ using \cite[(8.66)]{2005_David_Fundamentals}
{\setlength\abovedisplayskip{5pt}
\setlength\belowdisplayskip{4pt}
\begin{align}\label{eq:beam_opt}
	\hspace{-2mm}\mathbf w_{i,j}^{\star} & = \underset{\left\|\mathbf w_{i,j}\right\|^2 = 1}{\rm arg \max} \hspace{1.5mm}\frac{p_{i,j}\left|\mathbf w_{i,j}^H\mathbf a_{i,i,j}\right|^2}{\sum_{k=1,k\neq i}^{j-1}p_{k,j}\left|\mathbf w_{i,j}^H\mathbf a_{k,i,j}\right|^2 + \sigma_i^2\left\|\mathbf w_{i,j} \right\|^2} \nonumber\\
	& = \frac{\left( \sum_{k=1,k\neq i}^{j-1}p_{k,j}\mathbf a_{k,i,j}\mathbf a_{k,i,j}^H + \sigma_i^2\mathbf I_M\right)^{-1}\mathbf a_{i,i,j}}{\left\|\left( \sum_{k=1,k\neq i}^{j-1}p_{k,j}\mathbf a_{k,i,j}\mathbf a_{k,i,j}^H + \sigma_i^2\mathbf I_M\right)^{-1}\mathbf a_{i,i,j}\right\|}. 
\end{align}}%

\subsubsection{Optimizing $\left\lbrace \{\delta_j\}, \{\mathbf S_{i,j}\}, \{p_{k,j}\}\right\rbrace$}
For given $\{\mathbf w_{i,j}\}$ and $\{\mathbf v_j\}$, the remaining variables can be optimized by solving
{\setlength\abovedisplayskip{5pt}
\setlength\belowdisplayskip{4pt}
\begin{align}\label{P1-sub2}
\underset{\substack{\{\delta_j \geq 0\}, \{\mathbf S_{i,j} \succeq \mathbf 0\},\\ \{p_{k,j} \geq 0\}}}{\max} \hspace{2mm} \sum_{i=1}^K\sum_{j=i+1}^{K+1}\delta_jr_{i,j}	 \hspace{4mm} \text{s.t.} \hspace{2mm} \eqref{P1_cons:b}-\eqref{P1_cons:d},
\end{align}}%
where $r_{i,j} \triangleq \log_2\left( 1 + \frac{p_{i,j}b_{i,i,j}}{\sum_{k=1,k\neq i}^{j-1}p_{k,j}b_{k,i,j} + \sigma_i^2}\right)$ with $b_{k,i,j} \triangleq \left|\mathbf w_{i,j}^H\mathbf \Psi_{k,i}\mathbf v_j\right|^2$, $\forall k,i\in\mathcal K, j = i+1, \cdots, K+1$. Problem \eqref{P1-sub2} exhibits non-convexity stemming from the non-concave nature of the objective function and the non-convex constraint \eqref{P1_cons:b}. To facilitate the solution design, we reformulate the objective function as $\sum_{i=1}^K\sum_{j=i+1}^{K+1}\delta_j\left(f_{i,j} - g_{i,j}\right)$, where both $f_{i,j}$ and $g_{i,j}$ are concave functions defined by $f_{i,j} \triangleq \log_2\left(\sum_{k=1}^{j-1}p_{k,j}b_{k,i,j} + \sigma_i^2\right)$ and $g_{i,j} \triangleq \log_2\left(\sum_{k=1,k\neq i}^{j-1}p_{k,j}b_{k,i,j} + \sigma_i^2\right)$, respectively. The fact that the first-order Taylor expansion of any concave function at any point is its global upper bound motivates the utilization of the successive convex approximation (SCA) to tackle this issue. To be specific, with given local points $\{p_{k,j}^t\}$ in the $t$-th iteration, we have $g_{i,j}\left(\{p_{k,j}\}\right) \leq g_{i,j}\big(\{p_{k,j}^t\}\big) + \frac{\sum_{k=1,k\neq i}^{j-1}b_{k,i,j}\left( p_{k,j} - p_{k,j}^t\right)}{\left(\sum_{k=1,k\neq i}^{j-1}p_{k,j}^tb_{k,i,j} + \sigma_i^2\right)\ln2} \triangleq g_{i,j}^{\rm ub, \it t}\left(\{p_{k,j}\}\right)$. Then, a lower bound of the optimal value of problem \eqref{P1-sub2} can be acquired by solving   
\begin{subequations}\label{P1-sub2-sca}
	\setlength\abovedisplayskip{-0.1pt}
	\setlength\belowdisplayskip{6pt}
	\begin{eqnarray}
		&\hspace{-4mm}\underset{\substack{\{\delta_j \geq 0\}, \{\mathbf S_{i,j} \succeq \mathbf 0\},\\ \{p_{k,j} \geq 0\}}}{\max}& \hspace{-2mm}\sum_{i=1}^K\sum_{j=i+1}^{K+1}\delta_j\left( f_{i,j} - g_{i,j}^{\rm ub, \it t}\left(\{p_{k,j}\}\right) \right)  \\
		&\text{s.t.}& \hspace{-8mm} \eqref{P1_cons:b}-\eqref{P1_cons:d},
	\end{eqnarray}
\end{subequations}
which is still non-convex. Nevertheless, by applying the change of variables $\tilde p_{k,j} = \delta_jp_{k,j}$, $\forall k\in\mathcal K, j = k+1,\cdots,K+1$, and $\tilde{\mathbf S}_{i,j} = \delta_j\mathbf S_{i,j}$, $\forall i\in\mathcal K, j=1,\cdots,i$, problem \eqref{P1-sub2-sca} can be equivalently written as
\begin{subequations}\label{P1-sub1-sca-eqv}
	\setlength\abovedisplayskip{6pt}
	\setlength\belowdisplayskip{6pt}
	\begin{align}
		&\underset{\substack{\{\delta_j \geq 0\}, \{\tilde{\mathbf S}_{i,j} \succeq \mathbf 0\},\\ \{\tilde p_{k,j} \geq 0\}}}{\max} \hspace{2mm}\sum_{i=1}^K\sum_{j=i+1}^{K+1}R_{i,j}^{\rm lb, \it t}\\
		\text{s.t.} & \hspace{1mm}  \sum_{j=k+1}^{K+1}\tilde p_{k,j} \leq \sum_{j=1}^k\eta\delta_j\sum_{i = j}^K{\rm tr}\left(\mathbf A_{k,i,j}\tilde{\mathbf S}_{i,j}\right), \forall k \in\mathcal K, \\
		& \hspace{1mm} {\rm tr}\left(\tilde{\mathbf S}_{i,j}\right) \leq \delta_jP_i,\forall i\in\mathcal K, j = 1,\cdots, i, \hspace{2mm} \eqref{P1_cons:c}, 
	\end{align}
\end{subequations}
where $R_{i,j}^{\rm lb, \it t} \!\triangleq\! \delta_j\log_2\left(\frac{\sum_{k=1}^{j-1}\tilde p_{k,j}b_{k,i,j}}{\delta_j} + \sigma_i^2\right) - \delta_jg_{i,j}\big(\{p_{k,j}^t\}\big) \!-\! \frac{\sum_{k=1,k\neq i}^{j-1}b_{k,i,j}\left( \tilde p_{k,j} - \delta_jp_{k,j}^t\right) }{\left(\sum_{k=1,k\neq i}^{j-1}p_{k,j}^tb_{k,i,j} + \sigma_i^2\right)\ln2}$ and $\mathbf A_{k,i,j} \triangleq \mathbf \Psi_{k,i}^*\mathbf v_j^*\mathbf v_j^T\mathbf \Psi_{k,i}^T$. As a convex semi-definite program (SDP), problem \eqref{P1-sub1-sca-eqv} can be directly solved using standard solvers such as CVX. 

\subsubsection{Optimizing $\{\mathbf v_j\}$} Given other variables, we now focus on optimizing $\{\mathbf v_j\}$. We particularly note that, unlike $\{\mathbf v_j\}_{j\in\mathcal J\backslash\{1\}}$, which are involved in both the objective function and constraints, $\mathbf v_1$ only exists in the constraints. Therefore, $\mathbf v_1$ and $\hat{\mathbf v} \triangleq \{\mathbf v_j\}_{j\in\mathcal J\backslash\{1\}}$ are optimized sequentially using distinct approaches as follows. 

\paragraph{Optimizing $\mathbf v_1$} The subproblem with respect to (w.r.t.) $\mathbf v_1$ is a feasibility-check problem. To obtain a more efficient solution, we introduce the ``EH residual'' variables $\{\Delta_k\}$, and then arrive at the following problem  
\begin{subequations}\label{P1-sub31}
	\begin{eqnarray}
		&\underset{\mathbf v_1, \{\Delta_k\geq 0\}}{\max}& \sum_{k=1}^K \Delta_k \\
		&\text{s.t.}& \hspace{-6mm} \sum_{j=k+1}^{K+1}\delta_jp_{k,j} + \Delta_k \leq \eta\delta_1\sum_{i = 1}^K\mathbf v_1^T\mathbf C_{k,i,1}\mathbf v_1^* \nonumber\\
		&& \hspace{-6mm} + \sum_{j=2}^k\eta\delta_j\sum_{i = j}^K\mathbf v_j^T\mathbf C_{k,i,j}\mathbf v_j^*, \forall k \in\mathcal K, \label{P1-sub31_cons:b}\\
		&& \hspace{-6mm} \left|\left[\mathbf v_1\right]_n\right| \leq 1, \ \left[ \mathbf v_1\right]_{N+1} = 1, \ \forall n\in\mathcal N, \label{P1-sub31_cons:c}
	\end{eqnarray}
\end{subequations} 
where $\mathbf C_{k,i,j} \triangleq \mathbf \Psi_{k,i}^T\mathbf S_{i,j}\mathbf \Psi_{k,i}^*$, $\forall i,k\in\mathcal K, j = 1,\cdots,k$. The convex nature of $\mathbf v_1^T\mathbf C_{k,i,1}\mathbf v_1^*$ in \eqref{P1-sub31_cons:b} leads to the non-convexity of problem \eqref{P1-sub31} but permits the use of SCA. With a given local point $\mathbf v_1^t$ in the $t$-th iteration, $\mathbf v_1^T\mathbf C_{k,i,1}\mathbf v_1^*$ is lower bounded by its first-order Taylor expansion, denoted by $\mathcal E_{\mathbf C_{k,i,1}}^{\rm lb, \it t}(\mathbf v_1)\triangleq 2{\rm Re}\left\lbrace \left(\mathbf v_1^t\right)^T\mathbf C_{k,i,1}\mathbf v_1^*\right\rbrace - \left(\mathbf v_1^t\right)^T\mathbf C_{k,i,1}\left(\mathbf v_1^t\right)^*$. By replacing $\mathbf v_1^T\mathbf C_{k,i,1}\mathbf v_1^*$ with $\mathcal E_{\mathbf C_{k,i,1}}^{\rm lb, \it t}(\mathbf v_1)$, problem \eqref{P1-sub31} can be approximated as    
\begin{subequations}\label{P1-sub31-sca}
	\begin{eqnarray}
		&\underset{\mathbf v_1, \{\Delta_k\geq 0\}}{\max}& \sum_{k=1}^K \Delta_k \\
		&\text{s.t.}& \hspace{-7mm}  \sum_{j=k+1}^{K+1}\delta_jp_{k,j} + \Delta_k \leq \eta\delta_1\sum_{i = 1}^K\mathcal E_{\mathbf C_{k,i,1}}^{\rm lb, \it t}(\mathbf v_1) \nonumber\\
		&& \hspace{-7mm} + \sum_{j=2}^k\eta\delta_j\sum_{i = j}^K\mathbf v_j^T\mathbf C_{k,i,j}\mathbf v_j^*, \forall k \in\mathcal K, \hspace{2mm} \eqref{P1-sub31_cons:c}, 
	\end{eqnarray}
\end{subequations} 
which is a convex quadratically constrained program (QCP) and can be efficiently solved by CVX. 

\paragraph{Optimizing $\hat{\mathbf v}$} To proceed, we optimize $\hat{\mathbf v}$. By introducing slack variables $\{z_{i,j}\}$, the subproblem w.r.t. $\hat{\mathbf v}$ can be equivalently converted to
\begin{subequations}\label{P1-sub32}
	\begin{eqnarray}
		&\hspace{-5mm}\underset{\hat{\mathbf v},\{z_{i,j}\}}{\max}& \sum_{i=1}^K\sum_{j=i+1}^{K+1}\delta_j\log_2\left( 1 + z_{i,j}\right)  \\
		&\hspace{-1cm}\text{s.t.}& \hspace{-6mm} \frac{\mathbf v_j^H\mathbf B_{i,i,j}\mathbf v_j}{z_{i,j}} \geq \sum_{k=1,k\neq i}^{j-1}\mathbf v_j^H\mathbf B_{k,i,j}\mathbf v_j + \sigma_i^2, \nonumber\\
		&& \hspace{-6mm} \forall i\in\mathcal K, j = i+1,\cdots,K+1, \\
		&& \hspace{-6mm} \sum_{j=k+1}^{K+1}\delta_jp_{k,j} \leq \eta\delta_1\sum_{k=1}^K\mathbf v_1^T\mathbf C_{k,i,1}\mathbf v_1^*\nonumber\\
		&& \hspace{-6mm} + \sum_{j=2}^k\eta\delta_j\sum_{i = j}^K\mathbf v_j^T\mathbf C_{k,i,j}\mathbf v_j^*, \forall k \in\mathcal K, \\
		&& \hspace{-6mm} \left|\left[\mathbf v_j\right]_n\right| \leq 1, \ \left[ \mathbf v_j\right]_{N+1} = 1, \ \forall n\in\mathcal N, j \in\mathcal J\backslash\{1\}, \label{P1-sub32_cons:d}
	\end{eqnarray}
\end{subequations} 
where $\mathbf B_{k,i,j} \triangleq p_{k,j}\mathbf \Psi_{k,i}^H\mathbf w_{i,j}\mathbf w_{i,j}^H\mathbf \Psi_{k,i}$, $\forall k,i\in\mathcal K, j=i+1,\cdots,K+1$. The non-convexity of problem \eqref{P1-sub32} arises from the convex terms $\frac{\mathbf v_j^H\mathbf B_{i,i,j}\mathbf v_j}{z_{i,j}}$ and $\mathbf v_j^T\mathbf C_{k,i,j}\mathbf v_j^*$, which prompts us to substitute these convex terms by their first-order Taylor expansion-based affine under-estimators. In this way, the problem to be solved becomes the following convex QCP: \looseness=-1
\begin{subequations}\label{P1-sub32-sca}
	\setlength\abovedisplayskip{6pt}
	\setlength\belowdisplayskip{6pt}
	\begin{eqnarray}
		&\underset{\hat{\mathbf v_j},\{z_{i,j}\}}{\max}& \sum_{i=1}^K\sum_{j=i+1}^{K+1}\delta_j\log_2\left( 1 + z_{i,j}\right)  \\
		&\text{s.t.}& \hspace{-5mm}\mathcal F_{\mathbf B_{i,i,j}}^{\rm lb, \it t}(\mathbf v_j, z_{i,j}) \geq \sum_{k=1,k\neq i}^{j-1}\mathbf v_j^H\mathbf B_{k,i,j}\mathbf v_j + \sigma_i^2, \nonumber\\
		&& \hspace{-5mm} \forall i\in\mathcal K, j = i+1,\cdots,K+1, \\
		&& \hspace{-5mm} \sum_{j=k+1}^{K+1}\delta_jp_{k,j} \leq \eta\delta_1\sum_{k=1}^K\mathbf v_1^T\mathbf C_{k,i,1}\mathbf v_1^*\nonumber\\
		&& \hspace{-5mm} + \sum_{j=2}^k\eta\delta_j\sum_{i = j}^K\mathcal E_{\mathbf C_{k,i,j}}^{\rm lb, \it t}(\mathbf v_j), \forall k \in\mathcal K, \hspace{2mm}  \eqref{P1-sub32_cons:d},
	\end{eqnarray}
\end{subequations} 
where $\mathcal F_{\mathbf B_{i,i,j}}^{\rm lb, \it r}(\mathbf v_j, z_{i,j}) \!\triangleq\! \frac{2{\rm Re}\left\lbrace(\mathbf v_j^t)^H\mathbf B_{i,i,j}\mathbf v_j\right\rbrace }{z_{i,j}^t} \!-\! \frac{(\mathbf v_j^t)^H\mathbf B_{i,i,j}\mathbf v_j^t}{\left(z_{i,j}^t\right)^2}z_{i,j}$ and $\mathcal E_{\mathbf C_{k,i,j}}^{\rm lb, \it t}(\mathbf v_j)\triangleq 2{\rm Re}\left\lbrace (\mathbf v_j^t)^T\mathbf C_{k,i,j}\mathbf v_j^*\right\rbrace - (\mathbf v_j^t)^T\mathbf C_{k,i,j}(\mathbf v_j^t)^*$. 

\vspace{1mm}
\subsubsection{Convergence and Complexity Analysis}
It can be proved as in \cite{2018_Qingqing_Multi-UAV} that the proposed algorithm can generate a non-decreasing sequence of objective values of (P1) by alternately optimizing $\{\mathbf w_{i,j}\}$, $\left\lbrace \{\delta_j\}, \{\mathbf S_{i,j}\}, \{p_{k,j}\}\right\rbrace$, and $\{\mathbf v_j\}$. This, in conjunction with the fact that the sum throughput is upper bounded by a finite value, ensures the convergence of the proposed algorithm. In addition, the complexity of each iteration is dominated by solving the SDP in \eqref{P1-sub1-sca-eqv} and the QCPs in \eqref{P1-sub31-sca} and \eqref{P1-sub32-sca}. According to \cite[Theorem 3.12]{2010_Imre_SDR_complexity} and \cite[Table III]{2014_K.wang_complexity}, the complexity of solving problems \eqref{P1-sub1-sca-eqv},  \eqref{P1-sub31-sca}, and \eqref{P1-sub32-sca} is given by $\mathcal O\left(\ln(\frac{1}{\varepsilon}) \left( K^2M^{3.5} + K^4M^{2.5} + K^6M^{0.5}\right)\right)$, $\mathcal O\left(\ln(\frac{1}{\varepsilon}) \left(K^{1.5}N^{4.5} \!+\! K^{2.5}N^{3.5}  \!+\! K^{3.5}N^{2.5}\right)\right)$, and $\mathcal O\big(\ln(\frac{1}{\varepsilon}) \allowbreak \times \big(K^6N^{4.5} + K^7N^{3.5} + K^8N^{2.5}\big)\big)$, respectively, where $\varepsilon$ denotes the prescribed solution accuracy. Thus, the complexity of each iteration of the proposed algorithm is about $\mathcal O\big(\ln(\frac{1}{\varepsilon})\big(K^2M^{3.5} +  K^4M^{2.5} + K^6M^{0.5} + K^6N^{4.5} + K^7N^{3.5} + K^8N^{2.5}\big)\big)$. 

\vspace{-3mm}
\subsection{How To Solve (P2) and (P3)?}
By observing the formulations of (P1)-(P3), it is not hard to see that the algorithmic framework proposed for (P1) applies to (P2) and (P3). Due to the space limitation, the details of how to solve (P2) and (P3) are omitted here. Besides, similar to the complexity analysis in the previous subsection, the complexity of each iteration of the algorithm for (P2) is about $\mathcal O\big(\ln(\frac{1}{\varepsilon})\big( K^2M^{3.5} + K^4M^{2.5} + K^6M^{0.5} + K^6N^{4.5}\big)\big)$, and that for (P3) is about $\mathcal O\big(\ln(\frac{1}{\varepsilon})\big( KM^{3.5} + K^2M^{2.5} + K^3M^{0.5} + K^{1.5}N^{4.5} + K^{2.5}N^{3.5} + K^{3.5}N^{2.5}\big)\big)$ \cite{2010_Imre_SDR_complexity,2014_K.wang_complexity}.

\vspace{-3mm}
\section{Simulation Results}
\begin{figure}[!t]
	\setlength{\abovecaptionskip}{-0.1pt}
	\setlength{\belowcaptionskip}{-1pt}
	\vspace{-6mm}
	\centering
	\includegraphics[scale=0.6]{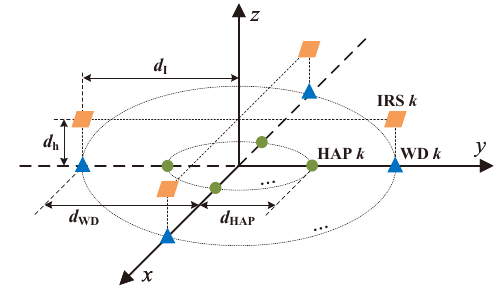}
	\caption{Simulation setup. HAP $k$ and WD $k$ are positioned in polar coordinates $\left(d_{\rm HAP},\frac{2\pi(k-1)}{K}, \frac{\pi}{2}\right)$ and $\left(d_{\rm WD},\frac{2\pi(k-1)}{K}, \frac{\pi}{2}\right)$ in meters (m), respectively. Each IRS is positioned $d_{\rm h}$ m directly above the y-axis (x-axis), with a vertical distance of $d_{\rm I}$ m from the z-axis.}\label{fig:simulation_setup}
    \vspace{-4mm}
\end{figure}

As depicted in Fig. \ref{fig:simulation_setup}, we consider a setup with $K = 4$ pairs of HAPs and WDs, and $L = K = 4$ IRSs, each equipped with $\frac{N}{4}$ elements. We adopt the distance-dependent path loss model and Rician fading channel model as detailed in \cite{2021_Xingquan_WPCN}. The path loss is $-30$ dB at the reference distance of $1$ m, with the path loss exponents of $2.2$ for the IRS-related links and $3.5$ for the direct links \cite{2019_Qingqing_Joint,2021_Xingquan_WPCN}. Moreover, the Rician factors of these two kinds of links are set to $3$ dB and $0$, respectively \cite{2022_Hua_WPCN}. If not otherwise specified, the following setups are utilized: $M = 2$, $N = 40$, $P_i = 33$ dBm, $\sigma_i^2 = -80$ dBm \cite{2019_Qingqing_Joint}, $\forall i\in\mathcal K$, $T = 1$ s, $\eta = 0.7$ \cite{2016_Hoon_WPCN}, $d_{\rm I} = d_{\rm WD} = 7$ m, and $d_{\rm h} = 2$ m. Each algorithm stops when the fractional increase of the objective value is below a convergence threshold $\epsilon = 10^{-3}$. 

\begin{figure}[!t]
	\vspace{-6mm}%
	\centering
	\setlength{\subfigcapskip}{-2pt}
	\setlength{\subfigbottomskip}{2pt}
	\subfigbottomskip = -6pt 
	\begin{subfigure}[]{\label{fig:rate_vs_dHAP}
			\includegraphics[scale=0.46]{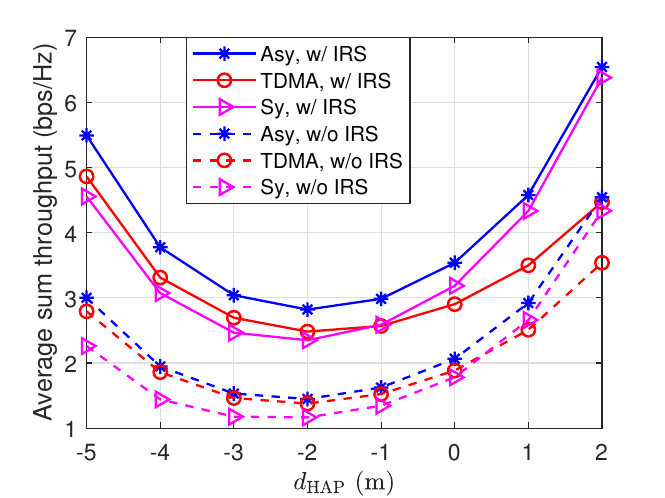}}
	\end{subfigure}
	\begin{subfigure}[]{\label{fig:energy_vs_dHAP}
			\includegraphics[scale=0.46]{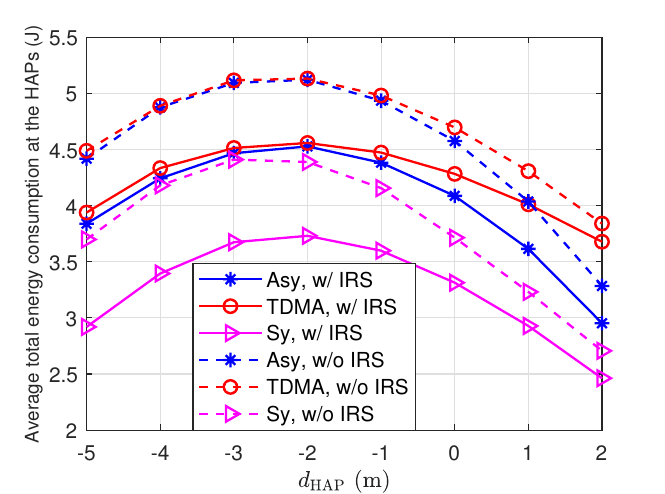}}
	\end{subfigure}
	\begin{subfigure}[]{\label{fig:rate_vs_N}
			\includegraphics[scale=0.46]{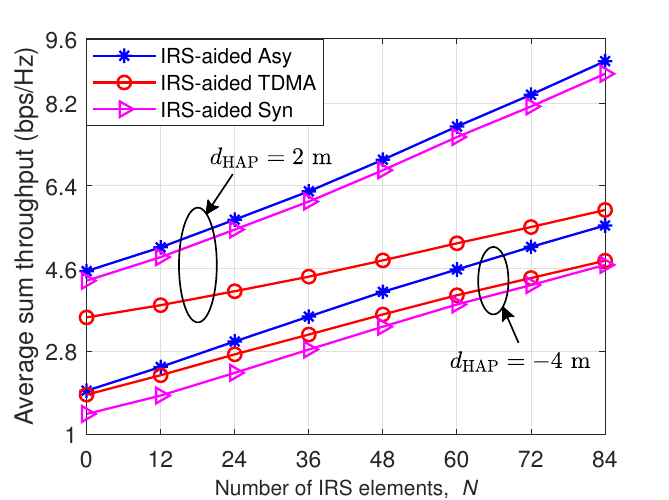}}
	\end{subfigure} \vspace{2mm}%
	\caption{(a) Average sum throughput versus $d_{\rm HAP}$; (b) Average total energy consumption at the HAPs versus $d_{\rm }$; (c) Average sum throughput versus $N$.}
	\vspace{-4mm}
\end{figure}

In Fig. \ref{fig:rate_vs_dHAP}, we study the impact of cross-link channel power on the system performance in both cases with and without IRSs by varying the value of $d_{\rm HAP}$. Here, $d_{\rm HAP} < 0$ means that HAP $k$ is located at $\left(-d_{\rm HAP},\frac{2\pi(k-1)}{K} + \pi, \frac{\pi}{2}\right)$ in m. First, it is observed that when $d_{\rm HAP} \geq -2$ m and increases, each scheme exhibits enhanced sum throughput performance. This can be attributed to the increased channel power of the direct links and reflected links (if present), which benefits both DL WPT and UL WIT. Second, when $d_{\rm HAP} \leq -2$ m and decreases, the sum throughputs of all the schemes increase. This is because the stronger cross-link channel power, although unfavorable for UL WIT, is advantageous for DL WPT, and these schemes can balance between DL WPT and UL WIT by optimizing the resource allocation to achieve better performance. Third, as expected, the Asy scheme consistently outperforms its sub-schemes, TDMA and Syn. Nevertheless, the performance gap between the Asy and Syn schemes decreases with the increase of $d_{\rm HAP}$. The reason is that the growing $d_{\rm HAP}$ results in reduced cross-link channel power, weakening the advantage of the Asy scheme in mitigating cross-link interference for more effective UL WIT through asymmetric time allocation and time-varying IRS beamforming (in the case with IRSs). Fourth, the Syn scheme gradually gains the upper hand over the TDMA scheme as $d_{\rm HAP}$ becomes large. This is because the performance gain brought by the TDMA scheme's advantage of being free from cross-channel interference in UL WIT gradually decreases as $d_{\rm HAP}$ increases, ultimately dropping below the performance loss incurred due to inefficient time resource usage. \looseness=-1

Next, we plot in Fig. \ref{fig:energy_vs_dHAP} the total transmission energy consumption at the HAPs, denoted by $E_{\rm total}$, versus $d_{\rm HAP}$ for different schemes. Combining Figs. \ref{fig:rate_vs_dHAP} and Fig. \ref{fig:energy_vs_dHAP}, we find that introducing IRSs into wireless powered IFCs provides dual benefits: enhancing the system throughput while concomitantly reducing the total energy consumption at the HAPs. Besides, we note that the trends of the curves depicting $E_{\rm total}$ are opposite to those of the curves illustrating system sum throughputs. This is understandable since the value of $E_{\rm total}$ is proportional to the duration of DL WPT, and the system sum throughput is proportional to the duration of UL WIT, but the durations of DL WPT and UL WIT are inversely related. \looseness=-1

\begin{table*}[!t]
	\vspace{-4mm}
	\centering
	\caption{Average sum throughput and average running time}\label{Tab}
	\footnotesize
	\belowrulesep = 0pt
	\aboverulesep = 0pt
	\begin{tabular}{c|cccccc||cccccc}
		\toprule\rule{0pt}{8pt}
		\multirow{3}{*}{} & \multicolumn{6}{c||}{Average sum throughput (bps/Hz)}                                                                                       & \multicolumn{6}{c}{Average running time (seconds)}                                                                                        \\ \cmidrule{2-13}\rule{0pt}{8.5pt}%
		& \multicolumn{3}{c|}{$\epsilon = 10^{-3}$}                                                         & \multicolumn{3}{c||}{$\epsilon = 10^{-2}$}                                    & \multicolumn{3}{c|}{$\epsilon = 10^{-3}$}                                                         & \multicolumn{3}{c}{$\epsilon = 10^{-2}$}             \\ \cline{2-13}\rule{0pt}{8pt} 
		& \multicolumn{1}{c}{Asy} & \multicolumn{1}{c}{TDMA} & \multicolumn{1}{c|}{Sy} & \multicolumn{1}{c}{Asy} & \multicolumn{1}{c}{TDMA} & Sy & \multicolumn{1}{c}{Asy} & \multicolumn{1}{c}{TDMA} & \multicolumn{1}{c|}{Sy} & \multicolumn{1}{c}{Asy} & \multicolumn{1}{c}{TDMA} & Sy \\ \hline\rule{0pt}{8pt}%
		$N = 12$            & \multicolumn{1}{c}{2.46}    & \multicolumn{1}{c}{2.26}     & \multicolumn{1}{c|}{1.84}   & \multicolumn{1}{c}{2.41}    & \multicolumn{1}{c}{2.26}     & 1.82   & \multicolumn{1}{c}{40.36}    & \multicolumn{1}{c}{11.29}     & \multicolumn{1}{c|}{25.50}   & \multicolumn{1}{c}{23.64}    & \multicolumn{1}{c}{11.00}     &  16.67  \\ 
		$N = 48$           & \multicolumn{1}{c}{4.09}    & \multicolumn{1}{c}{3.60}     & \multicolumn{1}{c|}{3.33}   & \multicolumn{1}{c}{3.91}    & \multicolumn{1}{c}{3.59}     &  3.19  & \multicolumn{1}{c}{162.75}    & \multicolumn{1}{c}{34.08}     & \multicolumn{1}{c|}{86.55}   & \multicolumn{1}{c}{54.70}    & \multicolumn{1}{c}{27.66}     &  30.96  \\ 
		$N = 84$            & \multicolumn{1}{c}{5.52}    & \multicolumn{1}{c}{4.77}     & \multicolumn{1}{c|}{4.67}   & \multicolumn{1}{c}{5.19}    & \multicolumn{1}{c}{4.76}     &  4.36  & \multicolumn{1}{c}{381.21}    & \multicolumn{1}{c}{60.13}     & \multicolumn{1}{c|}{191.96}   & \multicolumn{1}{c}{79.03}    & \multicolumn{1}{c}{46.64}     &  48.82  \\ \bottomrule 
	\end{tabular} \vspace{-2mm}
\end{table*}

Fig. \ref{fig:rate_vs_N} shows the average sum throughput versus $N$. It is evident that the TDMA scheme lags behind the Asy and Syn schemes in terms of sum throughput gains with increasing $N$. This is mainly due to the Asy and Syn schemes making better use of time resources to enhance performance, while the TDMA scheme struggles with time resource wastage. Furthermore, Figs. \ref{fig:rate_vs_dHAP}-\ref{fig:rate_vs_N} reveal that although the Asy scheme consistently achieves the highest sum throughput, the TDMA and Syn schemes may be more suitable in certain scenarios. Specifically, the TDMA scheme is preferred when cross-link interference is dominant and $N$ is small, as its performance approaches that of the Asy scheme while being easier to implement. Conversely, the Syn scheme is favorable when cross-link interference is low, offering comparable sum throughput performance to the Asy scheme while consuming less energy and being more practical to implement. 

Table \ref{Tab} shows the average sum throughput and corresponding average running time of the proposed schemes for $d_{\rm HAP} = -4$ m,  $N \in \{12,48,84\}$, and $\epsilon\in\{10^{-3}, 10^{-2}\}$. Simulations were performed using MATLAB 2021b on a laptop equipped with a 2.4 GHz Intel(R) Core(TM) i7-13700H processor and 16 GB of RAM.  It is observed that when $\epsilon$ increases from $10^{-3}$ to $10^{-2}$, the decrease in achieved sum throughputs is minor, while the reduction in running time is significant. In practice, a commercial high-speed computing device can shorten the running time to the scale of milliseconds, which is suitable for most scenarios. Besides, indoor Machine-Type Communication (MTC) devices, such as smart home sensors, industrial automation sensors, medical monitoring devices, and retail inventory sensors, generally experience long channel coherence times because they are deployed in fixed positions within stable and controlled environments. It is reasonable to expect the sum of running time and $T$ to remain below the channel coherence time. Thus, our proposed algorithmic framework is practical for real-world applications. 

\begin{figure}[!t]
	\setlength{\abovecaptionskip}{-0.1pt}
	\setlength{\belowcaptionskip}{-1pt}
	\vspace{-1mm}
	\centering
	\includegraphics[scale=0.46]{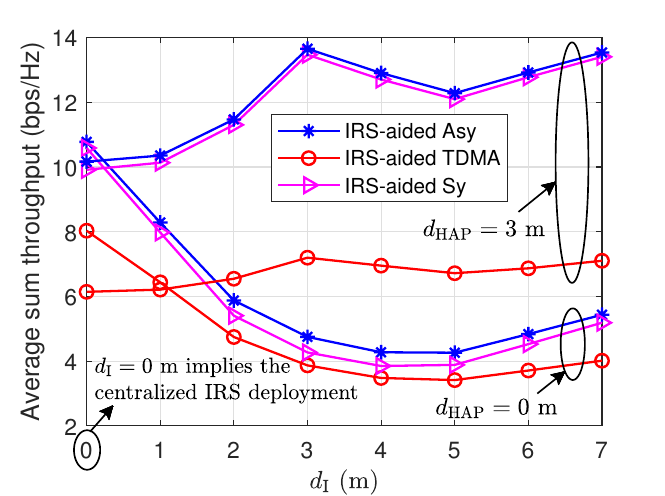}
	\caption{Average sum throughput versus $d_{\rm I}$.}
	\label{fig:rate_vas_dI}
	\vspace{-4.5mm}
\end{figure}

Fig. \ref{fig:rate_vas_dI} examines the impact of IRS deployment/location on the system performance when $d_{\rm h} = 1$ m. Here, $d_{\rm I} = 0$ m implies the centralized IRS deployment, and $d_{\rm I} \geq 1$ m corresponds to the distributed IRS deployment. We observe that when $d_{\rm HAP} = 3$ m (i.e., when the HAP-WD pairs are geographically dispersed), the distributed deployment outperforms the centralized deployment, while for $d_{\rm HAP} = 0$ m, the opposite holds. Furthermore, when the distributed deployment takes the upper hand in the case of $d_{\rm HAP} = 3$ m, deploying each IRS near its corresponding HAP or WD is optimal. In addition, different deployment strategies or locations do not affect the performance comparison results among the schemes.\looseness=-1

Finally, it is worth mentioning that we numerically observe different DL and UL IRS beamforming design results from our simulations. However, theoretically verifying the need for different IRS phase shifts in DL and UL for wireless powered IFCs remains an open issue, requiring further study.

\vspace{-2mm}
\section{Conclusion}
\vspace{-1mm}
In this paper, we proposed three transmission schemes, namely the IRS-aided Asy, TDMA, and Syn schemes, aiming to maximize the system sum throughput of a wireless powered IFC through resource allocation optimization. Despite the non-convexity of the three formulated problems, we proposed a general algorithmic framework applicable to each. Simulation results demonstrated the benefits of integrating IRSs into wireless powered IFCs in terms of both sum throughput performance and energy cost, offering insights into the most attractive scheme choices in certain scenarios. However, certain comparison results, particularly those related to energy cost, relied solely on simulations using our proposed suboptimal algorithmic framework. Deriving theoretical results to confirm or challenge these findings presents a challenging task that warrants further investigation. Besides, potential extensions of this work include scenarios with imperfect CSI, multi-antenna WDs, and multiple WDs within each cell. Particularly intriguing is comparing our proposed schemes to non-orthogonal multiple access (NOMA) in the context of multiple WDs within each cell. These aspects are left for future study. 

\vspace{-1.6mm}
\bibliographystyle{IEEEtran}
\bibliography{ref}

\end{document}